\documentclass[12pt]{article}

\title{Native vacancy energetics and association in TiN and CrN: chemical potential dependence, clustering, and surface effects}

\author{
\normalsize Barsha Bhattacharjee$^{a,b}$ and Emilia Olsson$^{a,b}$\\[0.4em]
\parbox{0.9\textwidth}{\centering
\footnotesize
$^{a}$ Institute of Theoretical Physics, University of Amsterdam, Science Park 904, 1098 XH Amsterdam, Netherlands\\
$^{b}$ Advanced Research Center for Nanolithography, Science Park 106, 1098 XG Amsterdam, Netherlands
}
}
\date{}
\usepackage{pdfpages}
\usepackage{graphicx} 
\usepackage[utf8]{inputenc}
\usepackage[english]{babel}
\usepackage{csquotes}
\usepackage{newunicodechar}
\newunicodechar{ }{\,}
\usepackage[version=4]{mhchem}
\usepackage{graphicx}
\usepackage{array}
\usepackage{lscape}
\usepackage{fixltx2e}
\usepackage{siunitx}
\usepackage{gensymb}
\usepackage{graphicx}
\usepackage[inkscapelatex=false]{svg}
\usepackage{dcolumn}
\usepackage{bm}
\usepackage{mhchem}
\usepackage{xcolor}
\usepackage{float}
\usepackage{underscore}
\usepackage{tabularx} 
\usepackage{amsmath}  
\usepackage{graphicx} 
\graphicspath{ {./images/} }
\usepackage{wrapfig}
\DeclareGraphicsExtensions{.png,.pdf}
\usepackage[margin=1in,letterpaper]{geometry} 
\usepackage[final]{hyperref} 
\hypersetup{
	colorlinks=true,       
	linkcolor=blue,        
	citecolor=blue,        
	filecolor=magenta,     
	urlcolor=blue         
}
\usepackage[style=numeric, sorting=none, backend=biber]{biblatex}  
\usepackage{array, booktabs, makecell,longtable, multirow}

\begin{document}
\maketitle {}

\begin{abstract}

Native vacancy populations in transition-metal nitride coatings reflect vacancy formation, association, and depth-dependent stability. Using density functional theory (DFT), we compare these effects in rocksalt TiN and CrN in bulk and at (001) surfaces, within phase-stability-constrained chemical potential ranges. In TiN, $V_\mathrm{N}$ is favoured under N-poor and $V_\mathrm{Ti}$ under N-rich conditions; in CrN, $V_\mathrm{N}$ remains preferred throughout its stability range. Association energies are referenced to corresponding isolated vacancies where positive values mean association lowers the energy and negative values favour separation. Mean bulk association energies for mixed metal-nitrogen pairs are $+0.132$ eV in TiN and $+1.053$ eV in CrN. At (001) surfaces, $V_\mathrm{N}$ in TiN prefers the surface and $V_\mathrm{Ti}$ the first subsurface layer. In CrN, $V_\mathrm{N}$ has similar formation energies at the surface and first subsurface layer, with a slight preference for the latter; $V_\mathrm{Cr}$ slightly prefers the second subsurface layer. Surface cluster energies are referenced to separated vacancies at their preferred depths. The compact $2V_\mathrm{Ti}+2V_\mathrm{N}$ cluster in TiN has favourable association in bulk ($+0.217$ eV per vacancy) but unfavourable association at the surface ($-0.346$ eV per vacancy). By contrast, $2V_\mathrm{Cr}+2V_\mathrm{N}$ in CrN remains stabilised, with its association energy decreasing from $+1.188$ to $+0.431$ eV per vacancy. CrN's stronger association occurs without larger structural distortions and is accompanied by distinct Cr-centred magnetic responses.

\end{abstract}

\section{Introduction}

The functional utility of transition metal nitrides (TMNs) in modern industrial applications, ranging from wear-resistant coatings \cite{Holleck1995MultilayerProtection, Senf1999WearCoatings} to diffusion barriers in microelectronic interconnects \cite{Hultman2000ThermalFilms}, is fundamentally governed by the presence and behaviour of intrinsic point defects \cite{Tsetseris2008TrappingStudy,Baben2012OriginFilms}. Their properties, however, are not determined solely by the ideal 1:1 compounds. Reactive deposition and annealing can produce deviations from stoichiometry \cite{Baben2012OriginFilms, Ponon2015EffectFilms, leFebvrier2022mathDefects}, while exchange with nitrogen reservoirs changes the chemical environment governing defect formation. Surfaces and interfaces further provide local environments in which defect stability can differ substantially from that in the bulk \cite{Zimmermann2009VacancySurface,Lee2012Environment-dependentVacancies}. Therefore, understanding such deviations from stoichiometry requires identifying which defects are energetically accessible and how the surrounding chemical and structural environments alter their stability.

In vacancies formed through exchange with external atomic reservoirs, the chemical environment is described by the elemental chemical potentials. The accessible range of these chemical potentials is constrained by the stability of the parent nitride relative to the elemental reference states and competing phases. Within these stability limits, variations in the nitrogen and transition metal chemical potentials modify the relative energetic costs of forming vacancies on the two sublattices. Consequently, any assessment of which vacancy type is favoured under different growth conditions must be based on a consistent treatment of the phase stability constraints. Although TMNs are often discussed in terms of nitrogen deficiency, non-stoichiometry can be accommodated through vacancies on either the nitrogen or metal sublattice, and the preferred mechanism will depend on the relative energetic cost of removing each species. Thus, determining which vacancy type is energetically favoured is essential for understanding defect populations, transport behaviour, and phase stability in these materials.  

Comparison of TiN and CrN is particularly relevant because they share the same rocksalt crystal structure while exhibiting distinct electronic and magnetic behaviour. TiN is metallic and remains cubic over a broad  stoichiometric range, with experimentally reported compositions of \(0.7 < x < 1.2\) and computational studies extending to \(x = 0.67\) \cite{Kang1999StructuralCompounds}. CrN also exhibits defect tolerance within its cubic rocksalt framework, but its structural stability is closely coupled to magnetic ordering, and the range over which cubic CrN remains stable depends on temperature, composition, and magnetic state \cite{Corliss1960AntiferromagneticCrN, Gall2002GrowthProperties, SubramanyaHerle1997SynthesisCr2S3,Gharavi2018MicrostructureFilms, Gharavi2021HighFilms}. These differences suggest that these two nitrides may not accommodate deviations from stoichiometry in the same way. Thus, the key concern becomes the choice of sublattice that preferentially loses atoms as the material departs from the ideal 1:1 stoichiometry. In the following, \(V_\mathrm{N}\) denotes a nitrogen vacancy and \(V_\mathrm{TM}\) a transition-metal vacancy, with \(V_\mathrm{Ti}\) and \(V_\mathrm{Cr}\) used for the corresponding material-specific metal vacancies. Multivacancy compositions are written additively, such that \(V_\mathrm{TM}+V_\mathrm{N}\) denotes a mixed transition metal–nitrogen divacancy.\par 
\par

Several components of this problem have previously been investigated using first-principles density functional theory (DFT). Tsetseris et al. (2007) calculated isolated nitrogen and metal vacancies, interstitials, and selected defect pairs in bulk TiN, HfN, and ZrN \cite{Tsetseris2007StructureNitrides}. For TiN, the studied \(2V_\mathrm{N}\) and \(2V_\mathrm{Ti}\) pairs were repulsive, demonstrating that the stability of an isolated defect does not by itself imply an aggregation tendency \cite{Tsetseris2007StructureNitrides}. Chemical-potential-dependent point-defect energetics has subsequently shown that $V_\mathrm{N}$ dominates Ti-rich, nitrogen-deficient Ti-N compounds, whereas $V_\mathrm{Ti}$ or nitrogen interstitials ($N_\mathrm{i}$) can become relevant toward N-rich compositions \cite{Connetable2024First-PrinciplesAlloys}. These studies establish the chemical potential dependence of dilute bulk defects, but do not resolve how the selected vacancies interact at finite concentration or near surfaces. \par

Prior work in CrN has primarily focused on nitrogen deficiency. Holec et al. (2014) compared several point-defect mechanisms in cubic CrN and identified $V_\mathrm{N}$ as the lowest-energy mechanism for reducing the N/Cr ratio. Their calculations further showed that $V_\mathrm{N}$ reduce Cr-to-N charge transfer and increase the metallic character of the defective material \cite{Holec2014ImpactCrN}. Zhang et al. (2013) identified ordered $V_\mathrm{N}$ on ({111}) planes in CrN and related increasing vacancy concentration to lattice contraction, structural distortion, and changes in ionicity and electronic structure \cite{Zhang2013InsightsCrN}. These results establish that $V_\mathrm{N}$ can occur in specific arrangements and interact with the CrN electronic structure. In our recent work, dilute intrinsic and extrinsic point defects were examined in CrN and CrN\(_2\), showing that the metallic CrN host comparatively screens isolated defects, such that their structural, electronic, and magnetic perturbations remain relatively modest \cite{Bhattacharjee2025ElucidatingNitride}. However, our previous study was restricted to the dilute-defect limit and did not address vacancy-vacancy association or near-surface segregation. The present work extends this understanding by examining how the relative stability of N and Cr vacancies evolves across the phase stability constrained chemical potential range, and by providing a configuration resolved comparison of nitrogen-only, chromium-only, and mixed vacancy associations.  

\par
Finite-concentration vacancy configurations have also been studied in other nitride systems. Cluster-expansion calculations for substoichiometric TiN$_x$ and VN$_x$ have identified vacancy-ordered structures and nitrogen-vacancy pair and triplet motifs, and examined their implications for elastic properties and hydrogen trapping and diffusion \cite{Cui2025VacancySteels}. More recent calculations of (Ti,V) ternary nitride precipitates have identified linear nitrogen-vacancy channels in V-rich domains \cite{Cui2026VacancySteels}. These studies focus on vacancy ordering within the nitrogen sublattice. Conversely, mixed $V_\mathrm{TM}+V_\mathrm{N}$ clusters have been shown to be experimentally relevant as Chen et al. (2025) observed Schottky-vacancy clusters in the MoN layers of MoN/TiN superlattices and demonstrated their role in intermixing and deformation \cite{Chen2025DirectSuperlattice}. Together, these studies show that vacancies can order within one sublattice or associate across both sublattices.

Surface studies provide a further, but incomplete, part of the picture. Low-index surface energies of TiN have been examined as functions of surface orientation and nitrogen chemical potential, with the non-polar (001) surface generally among the most stable terminations \cite{Wang2010SurfaceStudy}. \(V_\mathrm{N}\) formation and its layer dependence have also been studied for TiN surfaces in relation to work-function control \cite{Calzolari2020ControllingInvestigation}. For CrN(001), ideal and nitrogen-deficient surface structures have been compared, and selected N-deficient cubic terminations were found to be stable over parts of the chemical potential range \cite{Ponce-Perez2018StructuralStudies}. Experimentally, substoichiometric epitaxial TiN\(_x\)(001) films show pronounced composition-dependent changes in surface morphology and electrical resistivity, while epitaxial CrN(001) growth studies have demonstrated a strong sensitivity of surface morphology and physical properties to growth conditions \cite{Shin2004GrowthMgO001, Gall2002GrowthProperties}. These studies show that \(V_\mathrm{N}\) and surface stoichiometry affect stability and electronic properties but the question remains whether surfaces preserve or reorder the competition among \(V_\mathrm{N}\), \(V_\mathrm{Ti}\), and \(V_\mathrm{TM}+V_\mathrm{N}\), or how their association energies evolve from the outermost layer toward a bulk-like environment. Resolving this is particularly important for coating materials, where defect formation and redistribution in the near-surface region can directly influence growth, exchange with the surrounding environment, and the onset of degradation. Establishing how vacancy preferences change with depth is therefore essential for connecting surface defect chemistry to the stability and performance of the material under realistic conditions.

Thus, the relationship between vacancy selection and arrangement remains unresolved. While defect formation energy determines which defects are favoured in a given chemical environment, vacancy association energy determines how those defects organise once present. The former quantity depends on the elemental chemical potentials through exchange with external atomic reservoirs, whereas the latter compares configurations of identical compositions, for which the reservoir contributions cancel. Here, TiN and CrN are examined within a common framework linking phase stability, vacancy formation, association, and surface segregation. The chemically accessible N-rich and N-poor limits are defined from the stability of each nitride against competing phases, and defect formation energy is used to track changes in the preferred vacancy type across this range. Pair and higher-order association energies then establish whether vacancies remain dispersed or form same-sublattice or mixed TM-N arrangements. By considering vacancy pairs and higher-order clusters in addition to isolated defects, the analysis moves beyond the dilute-defect limit and probes local vacancy arrangements representative of finite defect concentrations. Finally, the same defect families are examined at the (001) surfaces to determine how vacancy stability and association evolve from the surface toward the bulk. This approach connects synthesis-relevant chemical conditions to vacancy chemistry and organisation across bulk and near-surface environments, while providing a framework that can be extended to other non-stoichiometric compounds.

\section{Methodology}

All calculations were performed using density functional theory (DFT) as implemented in the CP2K package, version 2024.3 \cite{Kuhne2020CP2K:Calculations}. The electronic structure was described using the Gaussian and plane-wave (GPW) method in the QUICKSTEP module \cite{VandeVondele2005Quickstep:Approach}, where the Kohn-Sham orbitals are expanded in atom-centred Gaussian basis functions and the electronic density is represented on an auxiliary plane-wave grid. Exchange-correlation effects were treated using the generalized gradient approximation (GGA) \cite{Langreth1983BeyondProperties} with the Perdew-Burke-Ernzerhof (PBE) functional \cite{Perdew1996GeneralizedSimple}. Core-valence interactions were described using Goedecker-Teter-Hutter pseudopotentials \cite{Goedecker1996SeparablePseudopotentials,Hartwigsen1998RelativisticRn} together with DZVP-MOLOPT-SR-GTH basis sets \cite{VandeVondele2007GaussianPhases}. Spin polarisation was included in all calculations. For CrN, the localized Cr \(3d\) states were treated using the Dudarev DFT+\(U\) approach with an effective Hubbard parameter of \(U_{\mathrm{eff}} = 3\) eV \cite{Anisimov1991BandiI/i,Dudarev1998Electron-energy-lossStudy}, consistent with previous work \cite{Zhou2014StructuralExperiment}. The auxiliary plane-wave cutoff and relative cutoff were set to 700 Ry and 60 Ry, respectively. These parameters were tested to ensure convergence of relative energies within 1 meV per atom. Electronic occupations were treated using Fermi-Dirac smearing with an electronic temperature of 300 K. Brillouin-zone sampling was restricted to the \(\Gamma\) point for all calculations.

The bulk phases of TiN and CrN were modelled in the B1 NaCl-type crystal structure. Bulk defect calculations were performed using \(4 \times 4 \times 4\) conventional supercells containing 512 atoms. The large supercell size reduces artificial interactions between periodic defect images and allows defect energetics to be evaluated at relatively dilute concentrations. For this cell size, a single vacancy corresponds to a concentration of 0.195\% relative to all atomic sites, or 0.391\% of the metal or nitrogen sub-lattice. The equilibrium lattice parameters of pristine TiN and CrN were obtained by fully relaxing both the cell vectors and the internal atomic positions. In defective supercells, the equilibrium cell shape and volume of the pristine structure were retained, while all internal atomic positions were allowed to relax. Electronic convergence was achieved using an SCF convergence criterion of \(10^{-6}\), and ionic relaxation was continued until the maximum residual force was below \(0.01\) eV/\AA{}. CrN was described using the Corliss-type antiferromagnetic structure \cite{Corliss1960AntiferromagneticCrN} (Corliss-AFM), in which ferromagnetically aligned Cr layers follow an alternating spin sequence along the \([110]\) direction. The relaxed TiN reference structure retained the cubic rocksalt geometry, with a lattice parameter of $a=4.247$~\AA{} and a nearest-neighbour Ti-N bond length of 2.124~\AA{}. The calculated lattice parameter is in close agreement with experimental values of approximately 4.24~\AA{}, differing by less than 0.2\% \cite{Straumanis1967BondingN}. The relaxed Corliss-AFM CrN structure retained the rocksalt connectivity with the characteristic low-temperature shear distortion. In the present supercell, the non-orthogonal angle is $\gamma=91.31^\circ$, corresponding to an acute pseudo-cubic shear angle of $88.69^\circ$. The volume-equivalent pseudo-cubic lattice parameter is $a_{\mathrm{pc}}=4.157$~\AA{}, and the nearest-neighbour Cr-N distances range from 2.074 to 2.084~\AA{}. Experimental low-temperature orthorhombic CrN lattice parameters correspond to a volume-equivalent pseudo-cubic lattice parameter of approximately 4.13~\AA{}, placing the present value within about 0.6\% of experiment \cite{Corliss1960AntiferromagneticCrN,Mayrhofer2008StructureCr1xAlxN}, whereas experimental cubic rocksalt CrN has a lattice parameter of $a=4.1467(8)$~\AA{} at 298~K, placing our value within approximately 0.25\% of the room-temperature cubic reference \cite{Aoki2002CrNMelt}. The calculated Cr-N distances are likewise consistent with the approximately 2.07~\AA{} nearest-neighbour separation expected from experimental CrN lattice parameters. Defects were introduced by removing metal atoms, nitrogen atoms, or combinations of both. To study defect clusters, different vacancy arrangements were considered by varying the relative separation, orientation, and local environment of the vacancies, allowing the effects of vacancy ordering and clustering on the formation energy to be assessed.

Defect formation energies ($E_{f}^{bulk}$) were calculated using \cite{Zhang1991ChemicalSelf-diffusion}:
\begin{equation}
\label{Edef}
E_{f}^{bulk} = E_{def}^{bulk} - E_{bulk} - \sum_{i} n_{i}\mu_{i},
\end{equation}
where $E_{def}^{bulk}$ and $E_{bulk}$ are the total energies of the defective and pristine cells, respectively. The term $n_i$ denotes the number of atoms of species $i$ added ($n_i > 0$) or removed ($n_i < 0$). The chemical potential $\mu_i$ defines the energy cost of exchanging atoms with the corresponding reservoir and therefore sets the growth-condition dependence of the defect formation energy. The chemical potential range was determined from the energetic stability of the host nitride with respect to elemental precipitation and competing transition-metal nitride phases. Chemical potentials were written as $\mu_i=\mu_i^0+\Delta\mu_i$, where $\mu_i^0$ is the elemental reference energy and $\Delta\mu_i \leq 0$ \cite{Buckeridge2014AutomatedCompounds}. For the stoichiometric host nitride, stability requires
\begin{equation}
\Delta\mu_{TM}+\Delta\mu_N=\Delta H_f(\mathrm{TMN}),
\end{equation}
where $\Delta H_f(\mathrm{TMN})$ is the formation enthalpy calculated relative to the elemental reference states of Ti, Cr, and $\frac{1}{2}N_2$. To capture limiting growth environments, defect formation energies were evaluated at both N-rich/TM-poor and N-poor/TM-rich limits \cite{Buckeridge2014AutomatedCompounds}. At the N-rich limit $(\Delta\mu_N=0)$ corresponding to equilibrium with the nitrogen reference; $\Delta\mu_{TM}$ is fixed by the host-stability condition. In the N-poor limit, the maximum allowed metal chemical potential was determined from the combined constraints imposed by elemental metal precipitation and competing TM$_x$N$_y$ phases. Where a metal-rich secondary nitride (such as $Ti_\mathrm{2}N$) imposed a stricter energetic bound, that phase defined the N-poor limit; otherwise, the elemental metal-rich boundary was retained. This procedure ensures that the selected chemical potential range remains within the stability window of the host nitride and does not favour precipitation of elemental reservoirs or formation of lower-energy competing nitride phases. The competing-phase constraints, numerical chemical potential limits, and Ti-N and Cr-N stability constructions are provided in Supplementary Table S1 and Figure~S1.

To determine whether vacancies that are already present tend to remain separated or associate into clusters, vacancy association energies (equivalent to the conventional vacancy binding energy), $E_{\mathrm{assoc}}^{\mathrm{bulk}}$, were evaluated relative to the isolated vacancies. The term association energy is used here to distinguish this process from defect formation since it describes the interaction between vacancies with a fixed overall composition. For a configuration containing $n$ $V_{\mathrm{TM}}$ and $m$ $V_{\mathrm{N}}$, the association energy was calculated as

\begin{equation}
    \label{Eassoc}
      E_{\mathrm{assoc}}^{\mathrm{bulk}}(n,m)=nE_f^{\mathrm{bulk}}(V_{\mathrm{TM}})+mE_f^{\mathrm{bulk}}(V_{\mathrm{N}})-E_f^{\mathrm{bulk}}(nV_{\mathrm{TM}}+mV_{\mathrm{N}})
\end{equation}

Here, $n,m \geq 0$ and $2 \leq n+m \leq 4$ for the multi-vacancy configurations considered in this study. The same expression applies to the TM-only ($m=0$), N-only ($n=0$), and mixed TM-N vacancy configurations. Since the associated configuration and its isolated-vacancy reference contain the same number of each atomic species, the chemical potential contributions to their formation energies are identical and cancel when the difference is taken. Consequently, $E_{\mathrm{assoc}}$ measures the interaction between vacancies, independent of the selected N-rich or N-poor chemical potential limit. The explicit expansion in terms of DFT total energies and cancellation of the chemical potential terms are provided in Supplementary Section S1. With the sign convention used here, a positive $E_{\mathrm{assoc}}$ indicates energetic stabilisation of the associated configuration relative to the isolated vacancies. A negative $E_{\mathrm{assoc}}$ indicates that the separated-vacancy reference is more energetically favourable. For the three- and four-vacancy configurations, the association energies are additionally reported per vacancy as $E_{\mathrm{assoc}}/(n+m)$ to facilitate a comparison between the configurations of different sizes. The pair association energies are reported as the total energies per pair because all pair configurations contain two vacancies.

Surface calculations utilized periodic slab models constructed from the optimised bulk structures. Low-index (001) surfaces were investigated for TiN and CrN. The slabs consisted of 11 atomic layers with a \(4 \times 4\) lateral expansion of the surface unit cell, and a vacuum region of 20 \AA{} was introduced along the surface normal to minimize interactions between periodic images. The Corliss-AFM ordering was retained in the CrN surface slabs. During structural relaxation, in-plane lattice parameters were fixed to the bulk equilibrium values while all atomic coordinates were fully relaxed. The surface free energy, $\gamma$, was defined as \cite{Qian1988StoichiometrySurfaces, Fiorentini1996ExtractingCalculations, Reuter2001CompositionPressure}:
\begin{equation}
    \gamma = \frac{1}{2A}(E_{slab} - NE_{bulk}) , 
\end{equation}
where $E_{slab}$ is the total energy of the relaxed slab, \textit{N} is the number of bulk formula units contained in the slab, and A is the surface area. These converged pristine surfaces served as the baseline for calculating depth-dependent defect stability, with vacancies introduced at surface, subsurface, and bulk-like central layers. Surface defect formation energies were evaluated using the same chemical potential formalism and the same elemental reference states and phase-stability limits as for the bulk defects. The pristine and defective bulk-supercell energies in the formation-energy expression were replaced by the pristine and defective slab energies, $E_{\mathrm{slab}}$ and $E_{\mathrm{def}}^{\mathrm{slab}}$, respectively. No separate surface chemical potentials were introduced. The same atomic reservoirs and chemical potential limits were used for the bulk and surface calculations, allowing the resulting defect formation energies to be compared on a consistent basis.

Structural changes induced by the defects were evaluated using root mean square (RMS) displacement as well as mean bond lengths. The RMS displacement was evaluated for atoms within 5 \AA{} of the defect centre. This cutoff includes the first few coordination shells around the vacancy complex in rocksalt TiN and CrN, while excluding atoms far from the defect that remain essentially bulk-like. The resulting $u_\mathrm{rms}^{5\AA}$ therefore provides a local measure of the defect-induced relaxation. For the surface structures, top-layer rumpling was defined as the difference between the mean $z$ coordinates of the transition-metal and nitrogen atoms in the outermost atomic layer, $r_0=\left|\langle z_{\mathrm{TM}}\rangle_0-\langle z_{\mathrm{N}}\rangle_0\right|$. Defect-induced changes in rumpling were evaluated relative to the pristine slab, $\Delta r_0=r_0^{\mathrm{def}}-r_0^{\mathrm{pris}}$.

\section{Results}

\subsection{Bulk vacancy energetics and association}

The bulk calculations first establish which vacancy sublattice is energetically preferred across the chemical potential range and then examine how vacancies interact when more than one defect is present, addressing the two questions raised in the Introduction.

\subsubsection{Vacancy-sublattice switching in TiN and its suppression in CrN} 

\begin{figure*}[h!]
        \centering \includegraphics[width=1\textwidth]{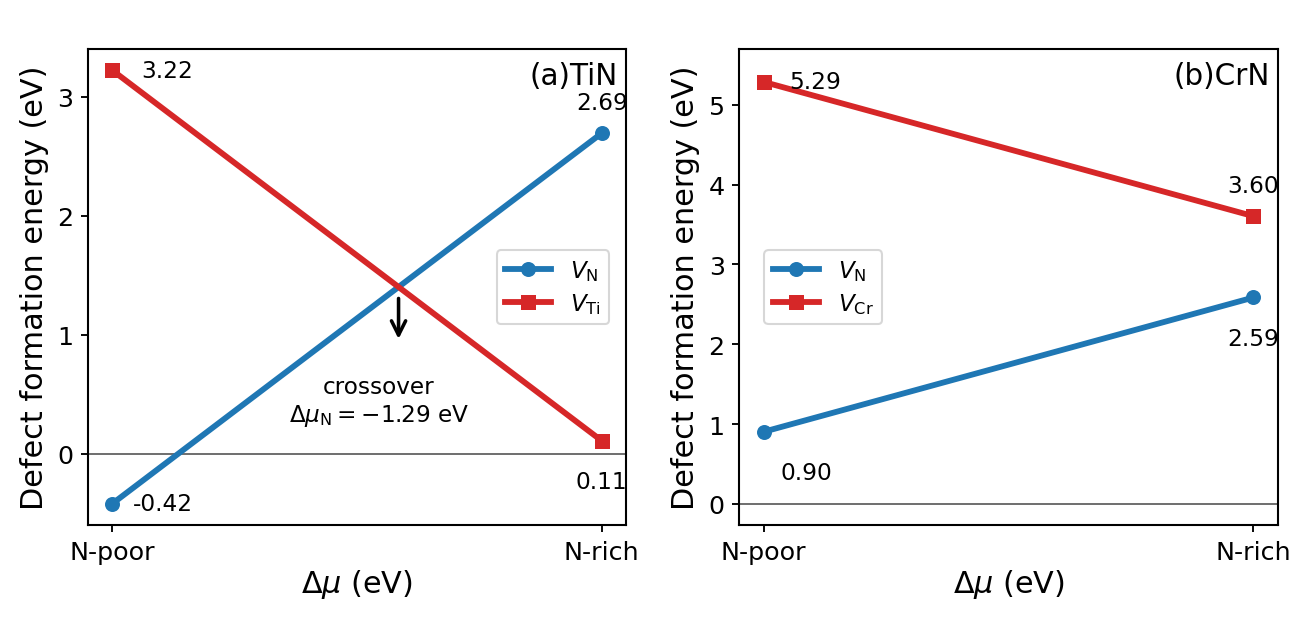}
        \caption{\label{figM1} $E_{f}^{bulk}$ of $V_\mathrm{TM}$ and $V_\mathrm{N}$ across the nitrogen chemical potential window considered for (a) TiN and (b) CrN.
        }
\end{figure*}

Figure~\ref{figM1} shows the bulk formation energies, $E_f^{\mathrm{bulk}}$, of $V_\mathrm{N}$ and the TM vacancy in (a) TiN and (b) CrN across their chemical potential ranges. These ranges are defined by the relevant elemental and competing phases, as summarised in Table~S1 and Figure~S1. At the N-poor/TM-rich limit, the stability boundary is set by Ti$_2$N for TiN and elemental Cr for CrN. A clear difference emerges between the two materials, with TiN undergoing a change in the preferred vacancy sublattice across its stability window, whereas CrN retains a preference for $V_\mathrm{N}$ throughout.

In TiN, the relative stability of $V_\mathrm{N}$ and $V_\mathrm{Ti}$ changes substantially with the chemical environment. At the N-poor/Ti-rich limit, $E_f^{\mathrm{bulk}}$ is $-0.418$~eV for $V_\mathrm{N}$ and $3.221$~eV for $V_\mathrm{Ti}$. Increasing $\Delta\mu_\mathrm{N}$ increases the formation energy of $V_\mathrm{N}$ and decreases that of $V_\mathrm{Ti}$, with the two formation energies becoming equal at $\Delta\mu_\mathrm{N}=-1.291$~eV. Above this crossover, $V_\mathrm{Ti}$ becomes energetically preferred. At the N-rich/Ti-poor limit, $E_f^{\mathrm{bulk}}$ is $0.110$~eV for $V_\mathrm{Ti}$ and $2.692$~eV for $V_\mathrm{N}$. TiN switches from an N-sublattice vacancy preference under N-poor conditions to a Ti-sublattice vacancy preference toward the N-rich limit.

The negative formation energy of $V_\mathrm{N}$ at the N-poor boundary indicates an energetic driving force for N removal from stoichiometric TiN under this limiting reservoir condition. At finite vacancy concentrations, defect-defect interactions become increasingly important, and the dilute-defect description becomes less applicable. The negative isolated-vacancy formation energy indicates a tendency toward N-deficient TiN close to the Ti$_2$N phase boundary. This chemical potential dependence is consistent with previous first-principles studies, which similarly predicted that $V_\mathrm{N}$ is favoured under Ti-rich/N-poor conditions and $V_\mathrm{Ti}$ becomes increasingly competitive toward N-rich conditions \cite{Connetable2024First-PrinciplesAlloys}. 

CrN shows a distinctly different response. At the N-poor/Cr-rich limit, $E_f^{\mathrm{bulk}}$ is approximately $0.904$~eV for $V_\mathrm{N}$ and $5.285$~eV for $V_\mathrm{Cr}$. Increasing $\Delta\mu_\mathrm{N}$ reduces the energetic separation between the two vacancy types but does not reverse their ordering. At the N-rich/Cr-poor limit, the formation energies are approximately $2.585$~eV for $V_\mathrm{N}$ and $3.604$~eV for $V_\mathrm{Cr}$. Thus, $V_\mathrm{N}$ remains the lower-energy isolated vacancy throughout the stability window, and no vacancy-sublattice crossover occurs.
This persistent preference for $V_\mathrm{N}$ is consistent with previous computational and experimental studies that identified N vacancies as the dominant defects associated with N deficiency in CrN \cite{Holec2014ImpactCrN,Zhang2013InsightsCrN}. The present calculations extend this picture by showing explicitly that $V_\mathrm{Cr}$ does not become the lower-energy isolated vacancy anywhere within the phase-stability-constrained chemical potential range considered here, including at the Cr-poor limit, where its formation energy is minimised.

Therefore, while TiN changes its preferred vacancy sublattice as the chemical environment varies, CrN retains a preference for nitrogen vacancies throughout the considered range. This indicates that sub-stoichiometry in CrN is expected to be accommodated through \(V_\mathrm{N}\). The following sections examine whether this contrast persists when multiple vacancies interact.

\subsubsection{Bulk pair vacancy interactions}

\begin{figure}[ht]
        \centering \includegraphics[width=0.80\textwidth]{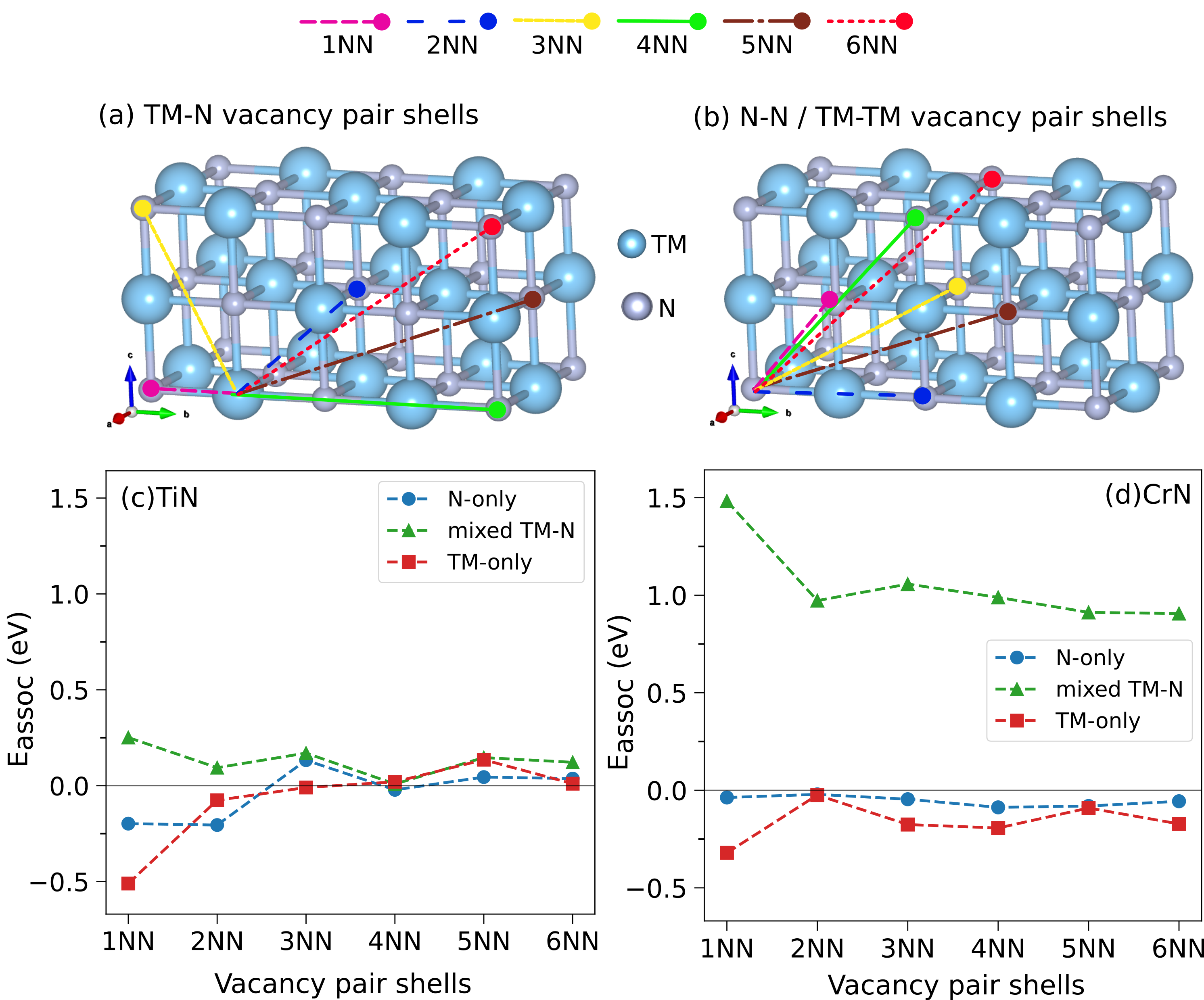}
        \caption{Vacancy-pair geometries and bulk association energies in TiN and CrN. (a) First- to sixth-nearest-neighbour (1NN-6NN) separations for mixed $V_{\mathrm{TM}}$-$V_{\mathrm{N}}$ pairs in the rocksalt lattice. (b) $V_{\mathrm{N}}$-$V_{\mathrm{N}}$ same-sublattice separations; the same geometrical shell sequence applies to $V_{\mathrm{TM}}$-$V_{\mathrm{TM}}$ pairs because both sublattices are fcc. Coloured dashed lines denote the respective neighbour shells. (c,d) Bulk vacancy-pair association energies, $E_{\mathrm{assoc}}$, for TiN and CrN, respectively, plotted as a function of neighbour shell for N-only, mixed TM-N, and metal-only vacancy pairs. Positive $E_{\mathrm{assoc}}$ indicates energetic stabilisation of the paired configuration relative to the isolated vacancies. Dashed connecting lines are guides to the eye only.
        \label{Fig2}
        }
\end{figure}

The isolated vacancy calculations show that TiN changes its preferred vacancy sublattice with chemical environment, whereas CrN retains a preference for $V_\mathrm{N}$. To determine how this picture changes when vacancies interact, the association energies of N-only, TM-only, and mixed TM-N pairs were compared from 1NN to 6NN. The mixed and same-sublattice pair geometries are shown in Figure ~\ref{Fig2}(a,b). Figure~\ref{Fig2}(c) shows that the pair interactions in TiN are small and strongly dependent on the pair geometry. The N-only pairs span $E_\mathrm{assoc}=-0.205$ to $+0.133$~eV, with a mean of $-0.035$~eV, while the Ti-only pairs span $-0.510$ to $+0.135$~eV, with a mean of $-0.072$~eV. All sampled mixed $V_\mathrm{Ti}$-$V_\mathrm{N}$ pairs have positive association energies, ranging from $+0.009$ to $+0.251$~eV, with a mean of $+0.132$~eV. The largest value occurs for the 1NN mixed pair, where $V_\mathrm{Ti}$ and $V_\mathrm{N}$ occupy directly neighbouring lattice sites separated by 2.124~\AA{}. The mean association energy of the 2NN-6NN mixed pairs is \(+0.108\)~eV, so the 1NN configuration is \(0.143\)~eV more strongly associated than the average of the 2NN-6NN configurations.

Structurally for the TiN pair configurations, the mean $u_\mathrm{rms}^{5\AA}$ values are $0.048$, $0.054$, and $0.062$~\AA{} for N-only, mixed, and Ti-only pairs, respectively. The mean Ti-N bond length changes are $-0.0063$, $-0.0065$, and $-0.0070$~\AA{}. The Ti-only pairs produce the largest average local displacements and bond contraction, with a mean association energy of $-0.072$~eV. The mixed pairs show smaller structural changes and a positive mean association energy of $+0.132$~eV. The strongest structural relaxation in TiN occurs for configurations with weaker vacancy association.

The CrN association energies plotted in Figure~\ref{Fig2}(d) show a much larger separation between the pair families. The N-only pairs span $E_\mathrm{assoc}=-0.088$ to $-0.021$~eV, with a mean of $-0.055$~eV, while the Cr-only pairs span $-0.321$ to $-0.024$~eV, with a mean of $-0.163$~eV. On the contrary, all six mixed $V_\mathrm{Cr}$-$V_\mathrm{N}$ pairs have positive association energies ranging from $+0.906$ to $+1.483$~eV, with a mean of $+1.053$~eV. The 1NN mixed pair, consisting of directly neighbouring Cr and N sites separated by 2.074~\AA{} (Figure~\ref{Fig2}(a)), has the largest association energy of $+1.483$~eV. The 2NN-6NN mixed pairs have a mean association energy of \(+0.967\)~eV, whereas the 1NN pair reaches \(+1.483\)~eV, with an additional stabilization of \(0.516\)~eV. Association remains above $+0.90$~eV at 6NN, where the vacancies are separated by 7.459~\AA{}, implying that strong mixed-vacancy association persists over the full range of separations considered. The enhanced stability of the 1NN mixed pair coincides with the compact geometry shown in Figure~\ref{Fig2}(a). The adjacent $V_\mathrm{Cr}$ and $V_\mathrm{N}$ form a single compact two-site defect region, producing a coordination environment distinct from that of the more widely separated configurations. The association energy of the 1NN configuration exceeds the mean of the 2NN-6NN configurations by $0.143$~eV in TiN and $0.516$~eV in CrN, indicating an additional energetic preference for direct TM-N vacancy adjacency. In CrN, the large positive association energies retained from 2NN to 6NN indicate that the mixed-vacancy interaction also extends well beyond the nearest-neighbour configuration.

Similar to TiN, the structural response of CrN also follows a different trend from the association energies. The mean $u_\mathrm{rms}^{5\AA}$ values for N-only, mixed, and Cr-only pairs are $0.022$, $0.046$, and $0.061$~\AA{}, respectively, while the mean Cr-N bond length changes are $-0.0014$, $-0.0044$, and $-0.0059$~\AA{}. Cr-only pairs produce the largest mean RMS displacements and bond contractions, with a mean association energy of $-0.163$~eV. The mixed pairs show intermediate structural changes and a mean association energy of $+1.053$~eV. The Cr magnetic response provides a further distinction between the pair families. Pristine CrN has a maximum local Cr moment of approximately $2.830~\mu_\mathrm{B}$. For the N-only pairs, the maximum Cr moment increases to $3.453$-$3.540~\mu_\mathrm{B}$, with a mean of $3.476~\mu_\mathrm{B}$. The mixed pairs have maximum Cr moments of $2.945$-$3.004~\mu_\mathrm{B}$, with a mean of $2.964~\mu_\mathrm{B}$, while the Cr-only pairs remain at $2.836$-$2.841~\mu_\mathrm{B}$. N-only pairs increase the mean maximum local Cr moment from $2.830~\mu_\mathrm{B}$ in pristine CrN to $3.476~\mu_\mathrm{B}$. For mixed $V_\mathrm{Cr}$-$V_\mathrm{N}$ pairs, the value is $2.964~\mu_\mathrm{B}$, only $0.134~\mu_\mathrm{B}$ above that of pristine CrN. Thus, the presence of $V_\mathrm{Cr}$ substantially reduces the change in the local moment associated with $V_\mathrm{N}$. This behaviour is consistent with the partial compensation of the local electronic perturbation associated with $V_\mathrm{N}$.

The pair calculations add an important feature to the isolated-vacancy picture. The formation energy of isolated $V_\mathrm{Cr}$ remains higher than $V_\mathrm{N}$ throughout the chemical potential range, whereas mixed $V_\mathrm{Cr}$-$V_\mathrm{N}$ pairs are strongly stabilised once both vacancy types are present. The mean mixed-pair association energy increases from $0.132$~eV in TiN to $1.053$~eV in CrN, a difference of $0.921$~eV. Thus, vacancy pairing provides a relatively small energetic gain in TiN and a much stronger driving force for mixed Schottky-type association in CrN.

At finite temperatures, these energetic differences may be reflected in the distribution and correlation of vacancies. The weaker association in TiN is consistent with a larger fraction of separated vacancies, while the CrN energetics favour correlated mixed $V_\mathrm{Cr}$-$V_\mathrm{N}$ configurations when both vacancy types are present. Vacancy ordering has been observed experimentally in N-deficient CrN \cite{Zhang2013InsightsCrN}, although the observations concern ordering of $V_\mathrm{N}$ on the N sublattice. The present calculations identify mixed-sublattice association as an additional energetically stable form of vacancy organisation in CrN. Such differences in vacancy distribution may influence diffusion, electronic and magnetic response, and mechanical behaviour. The following section examines whether these pair-interaction trends persist when vacancy complexes are extended to three- and four-vacancy clusters.

\subsubsection{Three- and four-vacancy clusters in bulk}

\begin{figure*}[h!]
        \centering \includegraphics[width=1\textwidth]{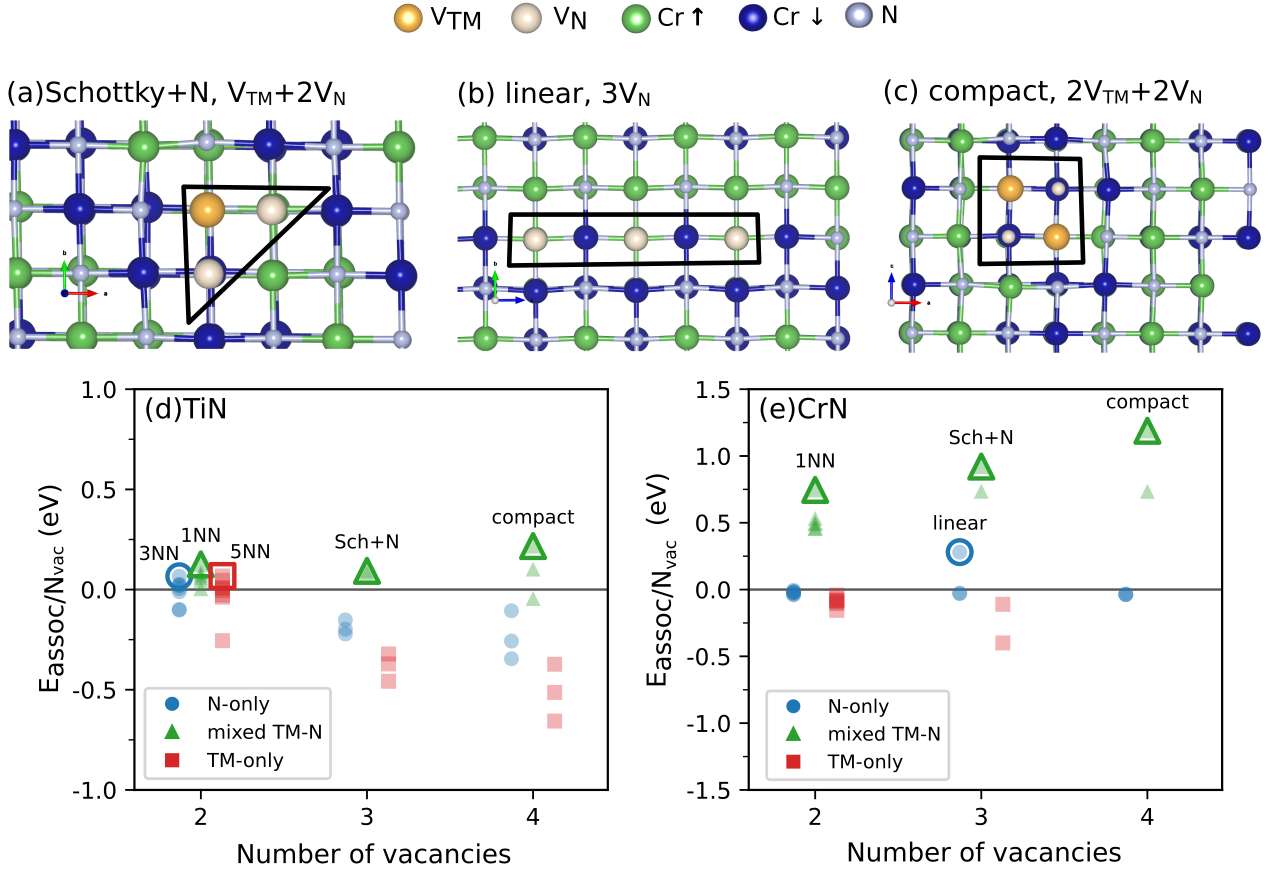}
        \caption{\label{figM3} Representative bulk vacancy-cluster geometries and association energies in TiN and CrN. (a-c) Higher-order vacancy motifs corresponding to configurations highlighted in the association-energy analysis: (a) a mixed three-vacancy Schottky-plus-N configuration, $V_{\mathrm{TM}}+2V_{\mathrm{N}}$ (Sch+N); (b) a linear three-nitrogen-vacancy configuration, $3V_{\mathrm{N}}$; and (c) a compact four-vacancy mixed configuration, $2V_{\mathrm{TM}}+2V_{\mathrm{N}}$. (d,e) Bulk $E_{\mathrm{assoc}}/N_\mathrm{vac}$ for TiN and CrN, respectively, as a function of the total number of vacancies. N-only, mixed TM-N, and metal-only clusters are shown separately. Faint filled symbols denote all converged configurations included in the analysis, whereas larger hollow symbols identify the most strongly stabilised representative configuration of each class at a given cluster size when $E_{\mathrm{assoc}}>0$. Positive association energies indicate energetic stabilisation of the clustered configuration relative to the isolated vacancies. Small horizontal offsets between the three vacancy classes are introduced solely for visual clarity and carry no physical meaning. Labels identify the highlighted representative motifs.
        }
        
\end{figure*}

The pair calculations showed that mixed TM-N association is weak in TiN but strong in CrN. Three- and four-vacancy configurations were examined to determine how these trends developed as more vacancies were introduced. Figure~\ref{figM3}(a-c) shows the clusters highlighted in the main analysis where (a) a mixed $V_\mathrm{TM}+2V_\mathrm{N}$ trimer, (b) a linear $3V_\mathrm{N}$ trimer, and (c) a compact $2V_\mathrm{TM}+2V_\mathrm{N}$ cluster. Additional bent, planar, segregated-zone, and separated-pair geometries are shown in Supplementary Figure S3. The association energies are reported per vacancy in Figure~\ref{figM3}(d,e), such that configurations containing two, three, and four vacancies can be compared on the same basis.

In TiN as shown in Figure~\ref{figM3}(d), adding a third vacancy does not strengthen the mixed-vacancy association observed for the 1NN pair. The 1NN $V_\mathrm{Ti}$-$V_\mathrm{N}$ pair has $E_\mathrm{assoc}/N_\mathrm{vac}=+0.126$~eV/vac, whereas the compact $V_\mathrm{Ti}+2V_\mathrm{N}$ and $2V_\mathrm{Ti}+V_\mathrm{N}$ trimers yield $+0.093$ and $+0.083$~eV/vac, respectively. The compact $2V_\mathrm{Ti}+2V_\mathrm{N}$ configuration is more strongly associated, reaching $+0.217$~eV/vac. The role of vacancy arrangement is especially clear when comparing the compact and separated $2V_\mathrm{Ti}+2V_\mathrm{N}$ configurations. The separated motif consists of two 1NN $V_\mathrm{Ti}$-$V_\mathrm{N}$ pairs, each with an internal separation of approximately 2.12~\AA{}, positioned so that the shortest distance between vacancies belonging to different pairs is approximately 12.38~\AA{} (Supplementary Figure S3). Its association energy is $+0.102$~eV/vac, which is close to the $+0.126$~eV/vac of the isolated 1NN mixed pair. Bringing the same four vacancies into the compact configuration raises $E_\mathrm{assoc}/N_\mathrm{vac}$ to $+0.217$~eV/vac, an increase of $0.115$~eV/vac. The structural response changes with this rearrangement. The separated-pair motif has $u_\mathrm{rms}^{5\AA}=0.030$~\AA{} and essentially no mean Ti-N bond-length change, whereas the compact $2V_\mathrm{Ti}+2V_\mathrm{N}$ cluster has $u_\mathrm{rms}^{5\AA}=0.073$~\AA{} and a mean Ti-N contraction of $-0.0078$~\AA{}. Therefore, the compact configuration allows a larger local rearrangement of the surrounding lattice. This greater local rearrangement accompanies the increase in association energy.

CrN retains the mixed-vacancy preference identified for the pairs [Figure~\ref{figM3}(e)]. The 1NN $V_\mathrm{Cr}$-$V_\mathrm{N}$ pair has $E_\mathrm{assoc}/N_\mathrm{vac}=+0.741$~eV/vac. The compact $V_\mathrm{Cr}+2V_\mathrm{N}$ and $2V_\mathrm{Cr}+V_\mathrm{N}$ trimers give $+0.921$ and $+0.736$~eV/vac, respectively, and the compact $2V_\mathrm{Cr}+2V_\mathrm{N}$ cluster reaches $+1.188$~eV/vac. The strong mixed association found for vacancy pairs persists as the complex grows, with the largest value occurring for the compact four-vacancy configuration. The separated and compact $2V_\mathrm{Cr}+2V_\mathrm{N}$ configurations isolate the effect of the vacancy arrangement because their compositions are identical. In the separated configuration, two 1NN mixed pairs retain internal $V_\mathrm{Cr}$-$V_\mathrm{N}$ separations of approximately 2.07~\AA{}, while the shortest separation between vacancies belonging to different pairs is approximately 13.40~\AA{} (Supplementary Figure S3). This configuration gives $E_\mathrm{assoc}/N_\mathrm{vac}=+0.736$~eV/vac, which is within $0.006$~eV/vac of the isolated 1NN-pair value. The compact cluster reaches $+1.188$~eV/vac, providing an additional $0.453$~eV/vac stabilisation when the two mixed pairs are brought into one compact vacancy complex. The associated structural change is also larger for the compact CrN cluster. The separated-pair configuration has $u_\mathrm{rms}^{5\AA}=0.018$~\AA{} and a mean Cr-N bond-length change of $+0.0003$~\AA{}, while the compact $2V_\mathrm{Cr}+2V_\mathrm{N}$ configuration has $u_\mathrm{rms}^{5\AA}=0.068$~\AA{} and a mean Cr-N contraction of $-0.0061$~\AA{}. The $0.453$~eV/vac increase in the association energy coincides with a greater ability of the neighbouring Cr-N framework to rearrange around a single compact vacancy region.

The local electronic response provides a second indication of why mixed configurations are stabilised in CrN. In pristine CrN, the mean local Cr Hirshfeld charge is approximately $-0.785\,e$. In the case of compact mixed trimers and tetramer, the corresponding values lie between $-0.794$ and $-0.761\,e$, within $0.024\,e$ of the pristine value. N-only three- and four-vacancy configurations shift the Cr charge to approximately $-0.741$ to $-0.687\,e$, yielding deviations of $0.044$-$0.098\,e$. Therefore, vacancies on both sublattices leave the local Cr charge closer to the pristine state than N-only vacancy clusters, which is consistent with partial electronic compensation. The magnetic response exhibited the same composition dependence. The maximum local Cr moment is $2.830~\mu_\mathrm{B}$ in pristine CrN. The magnetic moment is $2.918~\mu_\mathrm{B}$ for compact $2V_\mathrm{Cr}+V_\mathrm{N}$ and $2.926~\mu_\mathrm{B}$ for compact $2V_\mathrm{Cr}+2V_\mathrm{N}$, whereas the more N-rich $V_\mathrm{Cr}+2V_\mathrm{N}$ cluster reaches $3.363~\mu_\mathrm{B}$ and the N-only clusters span $3.473$-$3.733~\mu_\mathrm{B}$. Therefore, the mixed configurations maintain both the local charge and Cr moments closer to the pristine values. This indicates that electronic and magnetic compensation contributes to the preference for mixed Cr-N vacancy clusters alongside structural relaxation around the compact vacancy region.

The N-only CrN configurations show that vacancy number alone is insufficient to determine whether a cluster associates. The linear $3V_\mathrm{N}$ motif shown in Figure ~\ref{figM3}(b) has $E_\mathrm{assoc}/N_\mathrm{vac}=+0.280$~eV/vac, whereas the bent and compact $3V_\mathrm{N}$ configurations give $-0.030$ and $-0.027$~eV/vac, respectively. The investigated $4V_\mathrm{N}$ motifs remain between $-0.038$ and $-0.033$~eV/vac. The linear trimer also has a lower $u_\mathrm{rms}^{5\AA}$ of $0.020$~\AA{} than the compact and bent trimers, which give $0.025$ and $0.028$~\AA{}, respectively. Its maximum local Cr moment is $3.473~\mu_\mathrm{B}$, compared with $3.531$ and $3.733~\mu_\mathrm{B}$ for the compact and bent motifs, respectively. The positive association of the linear trimer reflects its particular vacancy arrangement rather than a general increase in the N-vacancy association with the cluster size. Preferred arrangements of this kind can provide local building blocks for vacancy ordering at higher defect concentrations \cite{Zhang2013InsightsCrN,Zhang2015Self-assemblyZrC1x,Ravi2010ClusterNitrides}.

All higher-order TM-only configurations with reported association energies have negative values. The Ti-only three- and four-vacancy motifs span $-0.656$ to $-0.321$~eV/vac, whereas the Cr-only motifs with reported energies span $-0.550$ to $-0.110$~eV/vac. Thus, none of the TM-only clusters with a reported association energy is stabilised relative to the isolated TM vacancies. The cluster results continue the contrast established by the pair calculations. In TiN, mixed trimers remain below $+0.10$~eV/vac, and only the compact $2V_\mathrm{Ti}+2V_\mathrm{N}$ configuration reaches $+0.217$~eV/vac. In CrN, mixed association persists from the 1NN pair at $+0.741$~eV/vac through the mixed trimers at $+0.736$ - $+0.921$~eV/vac and increases to $+1.188$~eV/vac for the compact four-vacancy cluster. CrN also supports the geometry-specific linear $3V_\mathrm{N}$ motif at $+0.280$~eV/vac. These association energies describe how an existing vacancy population can lower its energy by adopting a particular arrangement. TiN gains relatively little from clustering, except for the selected compact mixed configurations. CrN gains considerably more from bringing Cr and N vacancies into mixed complexes, while some N-only arrangements can also be stabilised when the geometry is favourable. This difference should affect the organisation of defects at finite concentrations and, in turn, the properties of the material.

Therefore, the bulk results establish two different modes of vacancy organisation. In TiN, the chemical environment controls the preferred vacancy sublattice, whereas the vacancy association remains comparatively weak. In CrN, isolated $V_\mathrm{N}$ remains energetically preferred, but mixed $V_\mathrm{Cr}$-$V_\mathrm{N}$ association becomes important once several defects are present. The next section examines how these bulk trends change near the (001) surface.

\subsection{Surface vacancy formation and association}

The bulk results establish the preferred vacancy sublattice and the strength of vacancy association in TiN and CrN. Near a surface, these trends can be modified by the reduced coordination and by the independent depth preference of each vacancy species. The following sections examine isolated vacancy stability as a function of depth, followed by pair association and higher-order clustering at the (001) surfaces.

\subsubsection{Isolated surface vacancies}

\begin{figure*}[h!]
        \centering \includegraphics[width=1\textwidth]{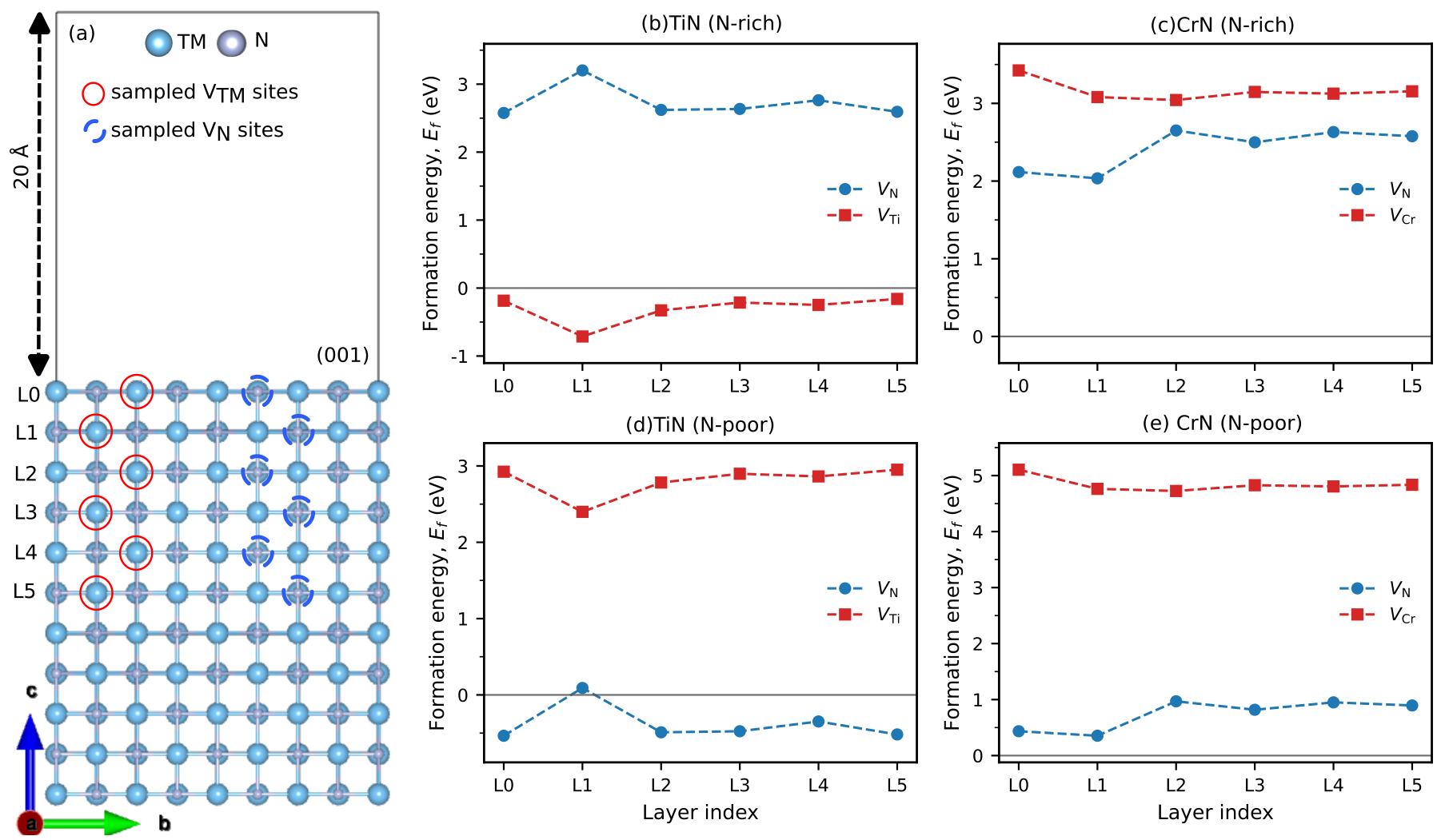}
        \caption{\label{figM4} Layer-resolved formation energies, $E_f$, of isolated vacancies at the TiN(001) and CrN(001) surfaces. (a) Slab geometry and layer indexing, with L0 denoting the outermost surface layer and increasing layer index indicating increasing depth into the slab. (b,c) Formation energies of $V_\mathrm{N}$ and the respective metal vacancy at the N-rich limit for TiN and CrN. (d,e) Analogous formation energies at the N-poor/TM-rich limit. The same phase stability constrained chemical potential limits are used as in the bulk calculations. Dashed lines connecting the discrete layer calculations are guides to the eye.
        }
\end{figure*}

Figure~\ref{figM4} shows the formation energies of isolated $V_\mathrm{N}$ and metal vacancies as a function of layer depth in TiN and CrN. Layer 0 denotes the outermost surface layer, and the layer index increases toward the centre of the slab, as illustrated in Figure~\ref{figM4}(a).  Changing the chemical potential shifts the formation energies of a given vacancy type by the same amount at every depth, and the variation with the layer index reflects the local surface environment.

In TiN, $V_\mathrm{N}$ is preferred over $V_\mathrm{Ti}$ at all sampled depths under N-poor conditions. The $V_\mathrm{N}$ formation energies at L0, L2, L3, and L5 are $-0.535$, $-0.490$, $-0.476$, and $-0.518$~eV, respectively, and differ by less than $0.06$~eV. L1 is the clear exception, with $E_f=+0.091$~eV, $0.609$~eV above the L5 value. The depth dependence of $V_\mathrm{N}$ arises mainly from an energetic penalty for the first subsurface site rather than from a gradual variation with distance from the surface. A surface N atom is already undercoordinated, whereas an N atom in L1 remains part of the subsurface Ti-N network while also bonding to the undercoordinated, outermost layer. Removing N from L1 is accompanied by minimal top-layer rumpling, from $0.175$ to $0.195$~\AA{} and changes of approximately $+0.13\%$ and $+0.08\%$ in the first two interlayer spacings. The local RMS displacement is also small, $0.035$~\AA{}. By L2 and L3, the formation energies are already close to the deepest-layer value. $V_\mathrm{Ti}$ shows the opposite response at L1. Under N-rich conditions, its formation energy decreases from $-0.160$~eV at L5 to $-0.712$~eV at L1, a stabilisation of $0.552$~eV. This configuration also produces a pronounced surface relaxation since the top-layer rumpling increases from $0.175$ to $0.331$~\AA{}, while the first two interlayer spacings contract by approximately $0.13\%$ and $0.23\%$. The local RMS displacement at L1 is $0.046$~\AA{}, smaller than the $0.069$~\AA{} obtained for a surface $V_\mathrm{Ti}$. Therefore, the L1 stabilisation coincides with a particular vertical rearrangement of the surface Ti and N sublattices rather than a larger overall displacement of the neighbouring atoms.

The chemical-potential-driven vacancy sublattice switching observed in bulk TiN persists throughout the slab. At the N-poor limit, $V_\mathrm{N}$ is lower in formation energy than $V_\mathrm{Ti}$ at all six depths, while the ordering is reversed at the N-rich limit. The crossover occurs between $\Delta\mu_\mathrm{N}=-1.38$ and $-1.96$~eV depending on depth, compared with $-1.291$~eV in the bulk. The largest shift occurs at L1, where the destabilisation of $V_\mathrm{N}$ and stabilisation of $V_\mathrm{Ti}$ move the crossover to $\Delta\mu_\mathrm{N}=-1.96$~eV. 

CrN retains its bulk preference for $V_\mathrm{N}$ at every sampled depth and at both chemical potential limits. The strongest near-surface stabilisation occurs for $V_\mathrm{N}$ at L0 and L1, where the formation energies are $0.460$ and $0.541$~eV lower, respectively, than at L5. The L1 configuration increases the top-layer rumpling from $0.037$ to $0.161$~\AA{}, while the first two interlayer spacings change by only $+0.06\%$ and $+0.01\%$. The electronic and magnetic responses are also the strongest for $V_\mathrm{N}$. For Cr atoms within 5~\AA{} of the vacancy, the mean Hirshfeld charge changes from approximately $-0.768\,e$ at L5 to $-0.739\,e$ at L1 and $-0.718\,e$ at L0. The maximum local Cr moment reaches $3.706~\mu_\mathrm{B}$ for L1 $V_\mathrm{N}$, compared with $3.062~\mu_\mathrm{B}$ at L0 and $3.276~\mu_\mathrm{B}$ at L5. Thus, L1 stabilisation coincides with enhanced surface rumpling and a particularly strong local magnetic response. The Hirshfeld charge variations remain modest across the sampled depths, indicating that charge redistribution is only one component of the stabilisation mechanism. $V_\mathrm{Cr}$ shows a weaker depth dependence. Its formation energy is approximately $0.27$~eV higher at L0 than at L5 and reaches its lowest value at L2, which is $0.11$~eV below that at L5. The maximum local Cr moments remain within approximately $2.89$-$2.94~\mu_\mathrm{B}$ across the sampled layers, and the nearby Cr Hirshfeld charges approach $-0.80\,e$ away from the surface. An L1 $V_\mathrm{Cr}$ produces a larger surface rumpling than the L2 configuration despite not being the lowest-energy site, showing that the magnitude of structural distortion alone does not determine the depth preference. The deepest slab layer reproduces the bulk vacancy ordering, although its absolute formation energy is not identical to the bulk value. For $V_\mathrm{N}$, the L5-bulk difference, which is the deepest slab reference is approximately $0.10$~eV in TiN and $0.01$~eV in CrN, while the differences for $V_\mathrm{Ti}$ and $V_\mathrm{Cr}$ are approximately $0.27$ and $0.45$~eV.

The surface calculations preserve the main bulk contrast. TiN retains its chemical potential-driven switch between N- and Ti-sublattice vacancies, with the first subsurface layer selectively destabilising $V_\mathrm{N}$ and stabilising $V_\mathrm{Ti}$. CrN retains its preference for $V_\mathrm{N}$ throughout the allowed chemical potential range, with additional stabilisation of N vacancies in the surface and first subsurface layers.

\subsubsection{Surface pair vacancy association}

\begin{figure*}[h!]
        \centering \includegraphics[width=\textwidth]{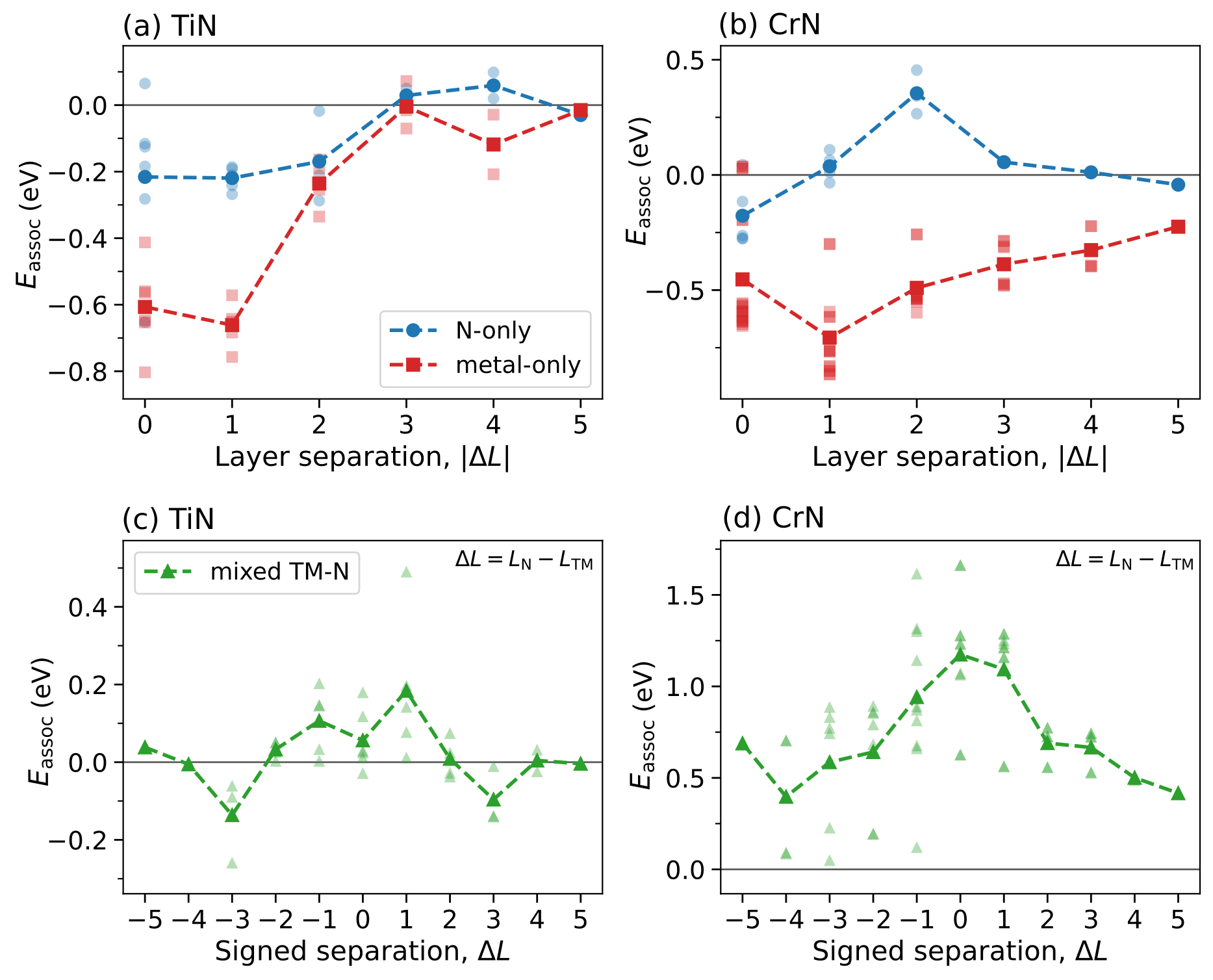}
        \caption{\label{figM5} Surface vacancy-pair association energies, $E_\mathrm{assoc}$, in TiN and CrN. (a,b) Same-sublattice N-N and TM-TM pairs as a function of the unsigned layer-index separation, $|\Delta L|=|L_1-L_2|$, for TiN and CrN, respectively. (c,d) Mixed TM-N pairs plotted using the signed separation $\Delta L=L_\mathrm{N}-L_\mathrm{TM}$. Positive $\Delta L$ places $V_\mathrm{TM}$ closer to the surface, negative $\Delta L$ places $V_\mathrm{N}$ closer to the surface, and $\Delta L=0$ denotes vacancies in the same layer. Faint symbols show individual configurations with different absolute layer positions, while larger symbols and dashed lines show the mean at each layer separation. Positive $E_\mathrm{assoc}$ indicates stabilization of the paired configuration relative to isolated vacancies occupying the same respective layers.
        }
\end{figure*}

To separate the vacancy-vacancy association from the independent depth preference of each defect, each surface pair was referenced to isolated vacancies occupying the same layers as the respective vacancies in the pair. A positive $E_\mathrm{assoc}$ indicates an energetic gain from the association at a fixed layer occupancy. In Figure~\ref{figM5}(a,b), the N-N and Ti-Ti/Cr-Cr pairs are shown, respectively. The horizontal axis is the unsigned layer-index separation, $|\Delta L|=|L_1-L_2|$, where $|\Delta L|=0$ denotes vacancies in the same layer. For the mixed Ti-N and Cr-N pairs, as shown in Figure~\ref{figM5}(c) and (d), the sign is retained using $\Delta L=L_\mathrm{N}-L_\mathrm{TM}$. Since the layer index increases into the slab, $\Delta L>0$ places $V_\mathrm{TM}$ closer to the surface, $\Delta L<0$ places $V_\mathrm{N}$ closer to the surface, and $\Delta L=0$ denotes a same-layer pair. Several configurations occur at a given $\Delta L$ because the same layer separation can be realised at different absolute depths. The faint symbols show the individually calculated configurations, while the larger connected symbols show the mean at each layer separation. The mean values are used only to highlight the overall dependence on layer separation. The individual configurations are retained because $\Delta L$ does not uniquely define the absolute depth, and in CrN, the magnetic Cr sublattice can additionally modify the interaction.

In TiN, the same sublattice vacancies show no tendency to associate (Fig. ~\ref{figM5}(a)). Across all sampled configurations, the mean association energies are $-0.138$~eV for N-N pairs and $-0.388$~eV for Ti-Ti pairs. N-N association becomes slightly positive for some larger layer separations, with mean values of $+0.029$ and $+0.059$~eV at $|\Delta L|=3$ and 4, respectively; however, no strong or systematic N-vacancy pairing tendency emerges. Ti-Ti shows an aversion to association, particularly for vacancies in the same or adjacent layers, where the mean association energies are $-0.607$ and $-0.661$~eV, respectively. Mixed $V_\mathrm{Ti}$-$V_\mathrm{N}$ pairs behave differently (Fig.~\ref{figM5}(c)), although the interaction remains strongly configuration dependent. Same-layer pairs have a mean $E_\mathrm{assoc}$ of $+0.056$~eV. For adjacent layers, the mean increases to $+0.184$~eV when $V_\mathrm{Ti}$ is closer to the surface ($\Delta L=+1$), compared with $+0.107$~eV for the reverse orientation ($\Delta L=-1$). The strongest calculated TiN mixed pair, with $V_\mathrm{Ti}$ in L0 and $V_\mathrm{N}$ in L1, has $E_\mathrm{assoc}=+0.491$~eV. Reversing the ordering to $V_\mathrm{N}$ in L0 and $V_\mathrm{Ti}$ in L1 gives only $+0.003$~eV. The latter arrangement closely matches the independently preferred isolated vacancy depths identified above. Thus, a strong local pair interaction can occur for a particular near-surface geometry without necessarily coinciding with the arrangement selected by isolated vacancy energies. Beyond the adjacent layers, the  mean Ti-N interaction rapidly approaches zero and can become slightly negative, showing that the surface TiN interaction is short-ranged and sensitive to the layer arrangement.

With regard to the structural response, the mean local RMS displacement within 5~\AA{} is $0.047$~\AA{} for N-N pairs, $0.052$~\AA{} for mixed Ti-N pairs, and $0.069$~\AA{} for Ti-Ti pairs. Thus, the Ti-Ti pairs produce the largest structural response, despite having the lowest affinity to associate. Similarly, the mean local Ti-N bond length changes are approximately $-0.004$, $-0.004$, and $-0.006$~\AA{} for N-N, Ti-N, and Ti-Ti pairs, respectively.  This implies that the energetic preference for the selected mixed configurations cannot be attributed to the magnitude of atomic relaxation.

The surface results are consistent with the weak association found in bulk TiN, where the mean Ti-N pair association energy is $+0.132$~eV compared to $+0.036$~eV over the sampled surface configurations. Individual near-surface arrangements can nevertheless be more strongly associated than the sampled bulk pairs, demonstrating that the surface broadens the range of possible pair interactions rather than introducing a general attraction. While the isolated-vacancy calculations show that the chemical environment determines which sublattice is preferentially depleted ($V_\mathrm{N}$ is stable toward N-poor conditions),  $V_\mathrm{Ti}$ becomes stable in the N-rich limit. The pair calculations show that vacancies on either sublattice have little energetic incentive to form same-sublattice pairs. At low vacancy concentrations, TiN non-stoichiometry is expected to be controlled primarily by vacancy formation and depth segregation, while mixed Ti-N associations become relevant only for selected local configurations.

CrN shows a qualitatively different pair-association landscape (Fig.~\ref{figM5}(b,d)). Similar to the Ti-Ti pairs, the Cr-Cr pairs also show no affinity towards association, with an overall mean $E_\mathrm{assoc}$ of approximately $-0.493$~eV. N-N interactions exhibit a distinct surface-induced feature; at $|\Delta L|=2$, all three sampled configurations tend to associate, with energies between $+0.266$ and $+0.455$~eV and a mean of $+0.355$~eV. This contrasts with bulk CrN, where the sampled N-N association energies remain slightly negative over the first six neighbour shells. Therefore, the surface can stabilise selected N-vacancy arrangements that are not favoured in the bulk. The dominant CrN interaction nevertheless remains mixed $V_\mathrm{Cr}$-$V_\mathrm{N}$ association [Fig.~\ref{figM5}(d)]. All sampled mixed surface configurations have positive $E_\mathrm{assoc}$, spanning $+0.051$ to $+1.664$~eV, with an overall mean of $+0.811$~eV. The same layer pairs are the strongest on average at $+1.174$~eV. Adjacent layer configurations also remain strongly associated, with mean values of $+1.093$~eV when $V_\mathrm{Cr}$ lies closer to the surface ($\Delta L=+1$) and $+0.942$~eV for the reverse ordering ($\Delta L=-1$). Association generally decreases as the vacancies become more widely separated in the layer index, producing a broad maximum at $\Delta L=0,\pm1$ in Fig. ~\ref{figM5}(d), but remains positive even for the largest sampled separations. The spread at a given $\Delta L$ shows that the absolute depth remains important.  

Therefore, the CrN surface preserves the strong mixed-vacancy tendency found in the bulk while making it more sensitive to the local environment. The bulk mixed pairs have a mean association energy of $+1.053$~eV over the first six neighbour shells, compared with $+0.811$~eV for the sampled surface configurations. The average stabilisation is consequently reduced by approximately $0.24$~eV at the surface. However, the strongest surface configuration, $+1.664$~eV, exceeds the bulk 1NN value of $+1.483$~eV. Therefore, the surface expands the interaction-energy distribution by introducing additional dependence on depth, orientation, and magnetic sublattice. The local structural, electronic, and magnetic response is consistent with this distinction. The mean local RMS displacement within 5~\AA{} is $0.028$~\AA{} for N-N pairs, $0.051$~\AA{} for mixed Cr-N pairs, and $0.073$~\AA{} for Cr-Cr pairs. As in TiN, the largest structural distortion occurs for the pair family which has negative association energies. In the magnetic response, N-N pairs produce the largest mean maximum local Cr moment of $3.51~\mu_\mathrm{B}$, compared with $3.03~\mu_\mathrm{B}$ for mixed Cr-N pairs and $2.87~\mu_\mathrm{B}$ for Cr-Cr pairs with the pristine value being $2.83~\mu_\mathrm{B}$. The enhanced moments around the N-N pairs are associated with the reduced N coordination of the neighbouring Cr atoms, which produces a stronger local electronic perturbation. When a Cr vacancy is associated with an N vacancy, this perturbation is partially reduced, and the largest local Cr moment shifts closer to the pristine value.  This trend is accompanied by a systematic change in the nearby Cr Hirshfeld charge, from approximately $-0.726\,e$ for N-N pairs to $-0.766\,e$ for mixed Cr-N pairs and $-0.800\,e$ for Cr-Cr pairs. Both the charge and magnetic responses show that the local electronic perturbation is strongest for N-vacancy-rich environments and is reduced when $V_\mathrm{Cr}$ is introduced into the defect complex.

The surface pair calculations answer the question posed in the Introduction differently for each material. TiN largely preserves its bulk defect hierarchy: chemical potential determines which vacancy sublattice is populated, surface depth controls where those vacancies preferentially reside, and pair association remains weak except for selected mixed near-surface geometries. CrN retains the strong bulk preference for mixed Cr-N  association, but the free surface introduces pronounced depth, orientation, and magnetic-sublattice dependence and additionally stabilises selected N-N arrangements. The dilute bulk hierarchy establishes the overall trend, while near-surface vacancy organisation is further shaped by depth, orientation, and the local surface environment.

\subsubsection{Three- and four-vacancy surface clusters}

\begin{figure*}[h!]
    \centering
    \includegraphics[width=\textwidth]{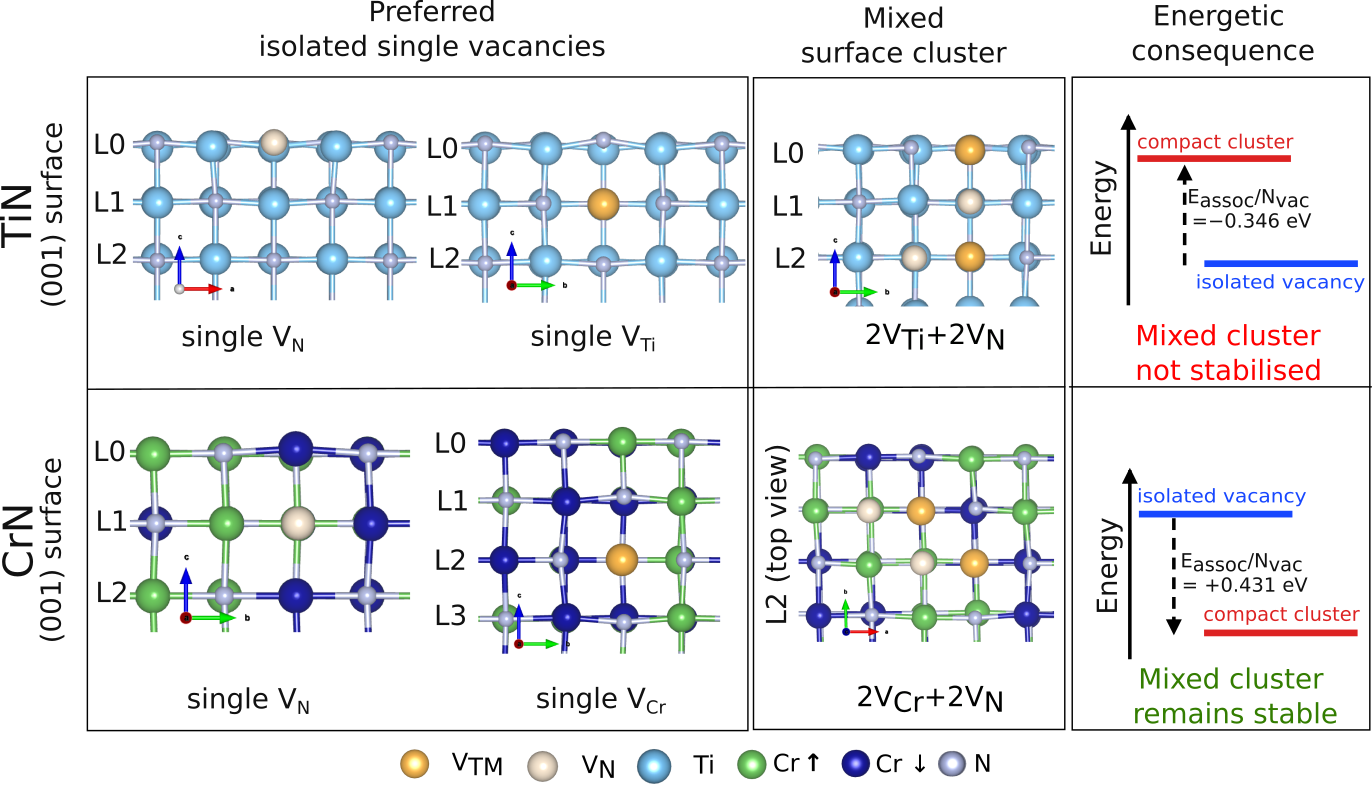}
    \caption{\label{figM6}
    Surface vacancy depth preference and representative higher-order vacancy association in TiN(001) and CrN(001). The TiN (top) and CrN (bottom) rows show the independently preferred sampled depths of the isolated vacancy species together with representative compact $2V_{\mathrm{TM}}+2V_{\mathrm{N}}$ surface clusters. For the cluster association energies, the separated reference state contains two isolated vacancies of each species at their respective preferred sampled depths; one representative structure of each isolated vacancy type is shown for clarity. The illustrated structures represent energetic reference states. Layer indices $L0,L1,\ldots$ are counted inward from the outermost surface layer. Positive $E_{\mathrm{assoc}}$ indicates that the compact cluster is stabilised relative to the preferred-depth separated-vacancy reference. The TiN motif is destabilised, with $E_{\mathrm{assoc}}/N_{\mathrm{vac}}=-0.346$~eV/vac, whereas the CrN motif remains stabilised, with $E_{\mathrm{assoc}}/N_{\mathrm{vac}}=+0.431$~eV/vac. Additional preferred-depth and pair-informed configurations are reported in the Supplementary Information to assess the sensitivity of these trends to the initial motif selection.
    }
\end{figure*}

The pair calculations establish whether two vacancies gain energy by associating at a fixed layer occupancy. To determine whether these pairwise interactions develop into compact multivacancy aggregates, three- and four-vacancy surface configurations were examined (Figure ~\ref{figM6}). The configurations were chosen for a targeted comparison with compact bulk motifs rather than an exhaustive sampling of all possible surface arrangements. To assess the sensitivity of these trends to the initial geometry-led sampling, additional motifs were constructed using the independently determined isolated vacancy depth preferences and strong layer-resolved pair interactions. For this comparison, the association energy was referenced to isolated vacancies occupying the lowest energy surface or subsurface layers. A positive $E_\mathrm{assoc}/N_\mathrm{vac}$ indicates that the compact aggregate is energetically stabilised, even after accounting for the energy gained when the constituent vacancies separately occupy their preferred depths.

In TiN, none of the investigated compact three- or four-vacancy configurations satisfies this condition. The $3V_\mathrm{N}$ and $4V_\mathrm{N}$ clusters have association energies of $-0.370$ and $-0.456$~eV per vacancy, respectively, while the $3V_\mathrm{Ti}$ and $4V_\mathrm{Ti}$ values are $-0.717$ and $-1.042$~eV per vacancy. Compact mixed configurations are also above the separated-vacancy reference, with $2V_\mathrm{Ti}+V_\mathrm{N}$, $V_\mathrm{Ti}+2V_\mathrm{N}$, and $2V_\mathrm{Ti}+2V_\mathrm{N}$ having $E_\mathrm{assoc}/N_\mathrm{vac}$ values of $-0.428$, $-0.513$, and $-0.346$~eV, respectively. The extension from vacancy pairs to the sampled three- and four-vacancy configurations does not strengthen the tendency to aggregate. Although select $V_\mathrm{Ti}$-$V_\mathrm{N}$ pairs show positive association energies for particular layer combinations, these configurations do not coincide with the preferred depths of the isolated vacancies. For the sampled larger TiN clusters, either the local association itself is insufficient to stabilise the aggregate, or any local energetic gain is outweighed by the cost of moving vacancies away from their preferred depths. The contrast with bulk TiN is particularly clear for the compact $2V_\mathrm{Ti}+2V_\mathrm{N}$ cluster, for which $E_\mathrm{assoc}/N_\mathrm{vac}$ changes from $+0.217$~eV in the bulk to $-0.346$~eV at the surface. Thus, the compact mixed clustering tendency observed for this bulk motif does not extend to the surface. The energetic penalty for compact Ti-vacancy-rich aggregation is accompanied by a substantially larger structural response than that of the analogous N-vacancy clusters. The local RMS displacement within 5~\AA{} increases from $0.053$ and $0.057$~\AA{} for $3V_\mathrm{N}$ and $4V_\mathrm{N}$ to $0.114$ and $0.132$~\AA{} for $3V_\mathrm{Ti}$ and $4V_\mathrm{Ti}$, respectively. The distinction is also visible in the surface rumpling, where the top-layer rumpling changes by only $0.011$~\AA{} for the compact N-vacancy clusters, but by $0.208$ and $0.281$~\AA{} for $3V_\mathrm{Ti}$ and $4V_\mathrm{Ti}$, respectively, indicating that the stronger energetic penalty for compact Ti-vacancy aggregation coincides with a much larger local and surface rearrangement.

CrN retains a stronger tendency toward mixed-vacancy aggregation. The compact N-only $3V_\mathrm{N}$ and $4V_\mathrm{N}$ surface clusters have $E_\mathrm{assoc}/N_\mathrm{vac}=-0.064$ and $-0.139$~eV per vacancy, respectively, and do not associate. In contrast, all investigated mixed Cr-N clusters have positive association energies.   Depending on the Cr magnetic sublattice, $2V_\mathrm{Cr}+V_\mathrm{N}$ gives $+0.317$-$+0.434$~eV per vacancy and $V_\mathrm{Cr}+2V_\mathrm{N}$ gives $+0.115$-$+0.199$~eV per vacancy, while $2V_\mathrm{Cr}+2V_\mathrm{N}$ gives $+0.431$~eV per vacancy. Mixed Cr-N association therefore survives beyond the pair level even when the competing depth preference of the isolated vacancies is included. The magnitude of this stabilisation is substantially lower than that of bulk CrN. The compact bulk $2V_\mathrm{Cr}+V_\mathrm{N}$ and $V_\mathrm{Cr}+2V_\mathrm{N}$ clusters have association energies of $+0.736$ and $+0.921$~eV per vacancy, respectively, compared with at most $+0.434$ and $+0.199$~eV per vacancy at the surface. For $2V_\mathrm{Cr}+2V_\mathrm{N}$, the association energy decreases from $+1.188$~eV per vacancy in bulk to $+0.431$~eV per vacancy at the surface, a reduction of $0.757$~eV per vacancy. Mixed Cr-N clustering remains energetically favourable at the surface, with a reduced association energy because the separated vacancies gain additional stabilisation from their preferred near-surface positions. Consistent with the pair calculations, the stronger mixed Cr-N vacancy association occurs without a larger structural distortion. The compact $3V_\mathrm{N}$ and $4V_\mathrm{N}$ clusters have $u_\mathrm{rms}^{5\AA}$ values of approximately $0.055$ and $0.056$~\AA{}, respectively, while the mixed clusters span $0.051$-$0.074$~\AA{}. These clusters can exhibit much stronger association without an accompanied larger structural displacement. In the magnetic response, the compact N-only clusters raise the maximum local Cr moment to approximately $4.00~\mu_\mathrm{B}$, whereas the mixed clusters generally reduce this response. The $2V_\mathrm{Cr}+V_\mathrm{N}$ configurations give maximum local Cr moments of  approximately $2.95$-$3.00~\mu_\mathrm{B}$ and the $2V_\mathrm{Cr}+2V_\mathrm{N}$ configurations give $3.20$-$3.27~\mu_\mathrm{B}$. The $V_\mathrm{Cr}+2V_\mathrm{N}$ configurations show a broader $3.05$-$3.67~\mu_\mathrm{B}$ range. Thus, the association of Cr and N vacancies partially moderates the strong local magnetic perturbation produced by N vacancies, consistent with an additional electronic and magnetic response accompanying the mixed Cr-N association. 

The three- and four-vacancy calculations extend the bulk-surface comparison beyond the dilute pair limit. In TiN, compact cluster growth is suppressed near the surface because vacancy-vacancy association must compete with the energetic preference of individual vacancies at different surface or subsurface depths. This differs from the bulk, where the association is governed primarily by the cluster composition and local geometry. Consequently, even though the selected mixed Ti-N pairs show positive layer-matched association energies, the sampled larger aggregates do not remain stabilised when referenced to separated vacancies occupying their preferred depths. In CrN, by contrast, the strong bulk preference for mixed Cr-N association persists from pairs to three- and four-vacancy clusters, although its magnitude is substantially reduced near the surface. Additional converged sensitivity clusters constructed from the isolated-vacancy depth preferences and strong layer-resolved pair interactions preserve this qualitative contrast. Mixed higher-order association remains unfavourable in TiN relative to preferred-depth isolated vacancies, whereas the sampled mixed Cr-N complexes remain positively associated. The preferred-depth $3V_\mathrm{N}$ motif in CrN also remains unfavourable (Supplementary Table~S6b). The resulting picture shows a material dependence  where the investigated TiN surface is better described by vacancies that remain dispersed and segregate to preferred depths, whereas CrN requires explicit consideration of mixed vacancy-vacancy correlations even in the near-surface region. This distinction is relevant because vacancy concentration and arrangement are known to influence the electronic and mechanical properties of substoichiometric transition metal nitrides \cite{Jhi2001VacancyNitrides}, while vacancy segregation and clustering near TiN surfaces and interfaces have also been implicated in the early stages of oxidation \cite{Zimmermann2009VacancySurface}. The present results show that near-surface vacancy configurations cannot be directly inferred from bulk association energies. Changes in the vacancy population may alter not only the overall non-stoichiometry of the TiN and CrN coatings but also whether vacancies remain separated, segregate to particular depths, or form mixed vacancy complexes near the surface.

\section{Conclusion}

This study provides a comparative first-principles description of native vacancy energetics in rocksalt TiN and CrN, connecting chemical-potential-dependent vacancy formation, vacancy association, higher-order clustering, and surface segregation within a common framework. Despite their similar crystal structures, the two nitrides accommodate vacancies differently. In TiN, the preferred isolated vacancy changes across the chemical potential range as $V_\mathrm{N}$ is favoured under N-poor/Ti-rich conditions and $V_\mathrm{Ti}$ is favoured under N-rich/Ti-poor conditions. In CrN, no similar crossover occurs, and $V_\mathrm{N}$ remains the lower-energy isolated vacancy throughout the chemical potential range.

The vacancy association further distinguishes between the two materials. Mixed TM-N vacancy pairs are the dominant attractive interaction in both TiN and CrN, but the association is substantially stronger in CrN. In bulk TiN, most pair and cluster interactions remain weak, although the compact $2V_\mathrm{Ti}+2V_\mathrm{N}$ motif is stabilised relative to isolated vacancies. CrN retains a strong mixed $V_\mathrm{Cr}$-$V_\mathrm{N}$ association from pairs to compact three- and four-vacancy clusters. The dependence of the association energy on cluster geometry, together with the stabilisation of selected N-vacancy motifs, shows that vacancy arrangement is important in addition to vacancy concentration. Such short-range organisation is relevant because the distribution of vacancies can modify the local electronic, magnetic, and mechanical responses of substoichiometric transition metal nitrides.

The surface introduces additional competition between vacancy-vacancy association and the independent depth preference of each vacancy. In TiN, $V_\mathrm{N}$ is favoured in the outermost layer, whereas $V_\mathrm{Ti}$ is most strongly stabilised in the first subsurface layer. Consequently, the local attraction found for selected mixed pairs is insufficient to stabilise the sampled compact higher-order aggregates relative to vacancies occupying their preferred depths. In CrN, $V_\mathrm{N}$ is strongly stabilised in the near-surface region while $V_\mathrm{Cr}$ shows only a weaker subsurface preference. This preferential stabilisation of N vacancies implies a tendency towards an N-deficient and, therefore, locally Cr-rich near-surface composition. Mixed $V_\mathrm{Cr}$-$V_\mathrm{N}$ association nevertheless remains strong enough to survive at the surface, particularly for same-layer and adjacent-layer configurations, although higher-order cluster stabilisation is reduced relative to the bulk.

Thus, across both bulk and surface environments, TiN is comparatively defect-dispersive, with vacancy populations strongly governed by the chemical potential and, near the surface, by segregation to preferred depths. CrN is more defect-associative, with $V_\mathrm{N}$ as the preferred isolated vacancy, while $V_\mathrm{Cr}$ and $V_\mathrm{N}$ show a strong tendency to associate into mixed complexes in both bulk and near-surface environments. The stronger association in CrN is not accompanied by a larger structural distortion, whereas substantial Cr-centred magnetic responses occur around the vacancy motifs, consistent with an important electronic and magnetic contribution to the TiN-CrN contrast. These results show that bulk vacancy energetics alone are insufficient to predict near-surface defect organisation and that both segregation and vacancy-vacancy correlations must be considered when relating coating composition to surface and interface behaviour. Such differences in near-surface vacancy populations and organisation may also influence subsequent oxidation. Future work will build on the present results by first examining oxygen incorporation at these vacancy configurations and then determining how the resulting oxidised defect environments affect hydrogen uptake.

\section*{Conflicts of interest}
The authors declare no conflicts of interest.

\section*{Data availability}
The data supporting this article are included in the
Supplementary Information and are available in Zenodo
at \url{https://doi.org/10.5281/zenodo.24495842}.

\section{Acknowledgements}
This work was conducted at the Advanced Research Center for Nanolithography, a public–private partnership between the University of Amsterdam (UvA), Vrije Universiteit Amsterdam (VU), Rijksuniversiteit Groningen (RUG), the Netherlands Organization for Scientific Research (NWO), and the semiconductor equipment manufacturer ASML. This work made use of the Dutch national e-infrastructure with the support of the SURF Cooperative using grant no.  EINF-12786, EINF-17838 and by NWO Domain Science under grant number 2025.051. The authors thank SURF (www.surf.nl) for the support in using the National Supercomputer Snellius. E.O. is grateful for a WISE Fellowship from the NWO and support via Holland High Tech through a public-private partnership in research and development within the Dutch top sector of High-Tech Systems and Materials (HTSM). 

\clearpage
\includepdf[pages=-]{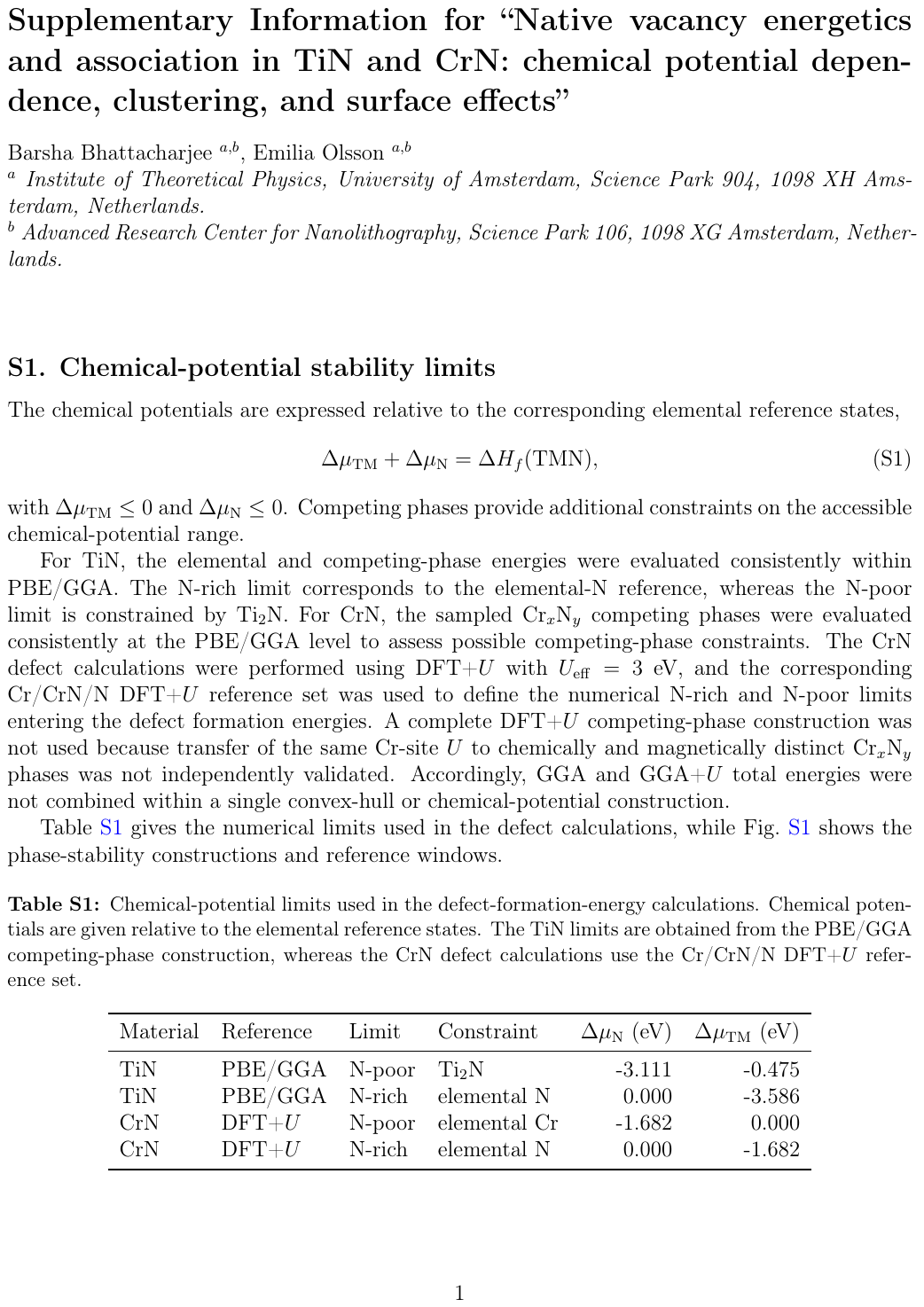}
\printbibliography
\end{document}